# Real-Symmetric Hamiltonian Enables Near-Linear Scaling for Fast Million-Atom Electronic Structure Computations

Zichong Zhang[1], Shuze Zhu*[1]

[1]Center for X-Mechanics, Department of Engineering Mechanics, Zhejiang University, Hangzhou 310000, China

*To whom correspondence should be addressed. E-mail: shuzezhu@zju.edu.cn

## Abstract

The exploration of quantum phenomena in mesoscale materials, such as moiré superlattices, is limited by the cubic-scaling cost of conventional electronic structure methods. Here, we introduce a scalable tight-binding framework that achieves near-linear scaling, enabling mesoscopic quantum simulations. By transforming the complex Hermitian Bloch Hamiltonian into an equivalent real-symmetric form, the method avoids dense diagonalization by combining sparse LDL decomposition with Sylvester's law of inertia for spectral slicing and global rank calibration. This formulation enables efficient band-structure calculations for large-scale systems, solving magic-angle twisted bilayer graphene in minutes on a standard laptop and extending to 1.5 million atoms within days on a single workstation. Applying this framework to ultra-low twist-angle structures with atomistic strain relaxation, we find robust isolated low-energy band clusters over several finite ultra-low-angle windows down to 0.09°. Our framework provides an efficient computational platform for studying quantum materials at experimentally relevant length scales and supports data-driven discovery in large-scale moiré systems.

## Introduction

The electronic properties of many modern quantum materials are increasingly shaped by physics that emerges at mesoscopic length scales. Representative examples include flat bands and correlated states in twisted van der Waals heterostructures [1–4], as well as topologically protected quantum spin Hall edge states [5–8]. Understanding these phenomena requires accurate electronic-structure descriptions of systems whose unit cells can contain hundreds of thousands to millions of atoms [4], a scale at which the computational cost becomes a major barrier.

While density functional theory (DFT) provides a first-principles foundation, standard orbital-based Kohn-Sham DFT calculations typically scale cubically with system size [9]. The high actual computational cost typically limits its application to systems containing $10^2 - 10^3$ atoms [10,11]. Tight-binding (TB) models therefore offer a practical balance between accuracy and efficiency and have therefore been widely used for larger systems [12–14]. Nevertheless, conventional TB calculations still rely on explicit diagonalization of the sparse Hamiltonian matrix, whose computational cost scales as $O(N^3)$ [14–17]. As a result, exact-diagonalization-based TB calculations are typically limited to unit cells with at most tens of thousands of orbitals [14,18], leaving a substantial scale gap to realistic low-angle moiré nanostructures. For example, in twisted bilayer graphene (TBG) with twist angles below $0.3°$ [19], the moiré period reaches tens of nanometers, so hundreds of thousands of atoms must be treated to capture electronic coupling and structural relaxation. The requirement for a large system is not solely computational but also physical: in low-angle TBG, moiré-scale lattice reconstruction produces localized vibrational and electronic modulations that must be resolved on the superlattice length scale [18]. Ultra-low-angle TBG therefore requires full commensurate supercells to consider moiré-scale relaxation and resolve electronic band structure [19,20]. Although several targeted eigensolvers [21] and DFT-based schemes have been developed to study moiré structures containing tens of thousands of atoms [22], extending these approaches to the million-atom regime remains a major challenge. For example, direct first-principles calculations of the magic-angle TBG moiré cell are already extremely demanding: ab initio calculations have been reported to require about 30-40 days per k point [23], while a fully relaxed plane-wave DFT study used thousands of CPU cores for about one month [24]. Even highly optimized first-principles implementations require supercomputing resources for a graphene system containing about $14000$ atoms [25].

In this work, we introduce a computational framework that overcomes the $O(N^3)$ bottleneck and enables tight-binding band-structure calculations at the million-atom scale. We demonstrate its capability by applying it to a series of TBG structures across an ultra-low twist-angle range, a regime previously inaccessible to ab initio or conventional TB approaches. Computationally, the proposed workflow turns $1.08°$ TBG band calculations into minutes-scale task on a standard laptop, while only requires several days for ultra-low-angle supercells containing up to 1.5 million atoms on a single workstation. We show that, when atomistic structural relaxation is included, isolated low-energy band clusters are manifested over several finite ultra-low-angle windows down to 0.09°. We further show that perpendicular magnetic fields enhance the isolation and flattening of the low-energy minibands, while Stone-Wales defects induce strongly location-dependent band perturbations that are most pronounced near AA regions. By bridging atomistic tight-binding accuracy and mesoscopic quantum length scales, our framework provides a scalable route for both fundamental studies and data-driven discovery of mesoscopic phenomena in complex materials [26–31].

## Computational Framework

The overall workflow of our method is summarized in Fig. 1 and applies to periodic systems described by a real-space tight-binding Hamiltonian. Starting from an atomic structure under periodic boundary conditions, we first construct a sparse Hermitian Hamiltonian using KD-tree-accelerated neighbor search [32], an essential step for treating large systems efficiently. We then determine the Fermi level at a reference wave vector through a combined spectral-window extraction and rank-calibration procedure. Shift-invert Arnoldi eigensolver [33,34] computes a small cluster of eigenvalues near a trial energy. Real-symmetric embedding of the complex Hermitian Hamiltonian enables sparse LDL factorization and Sylvester-inertia counting to identify the global ranks of energy spectrum without full diagonalization. For example, once the Fermi-level is determined, the near-Fermi bands are extracted along the Brillouin-zone path using the shift-invert Arnoldi method, with different high-symmetry points computed in parallel. Each step is described in detail below.

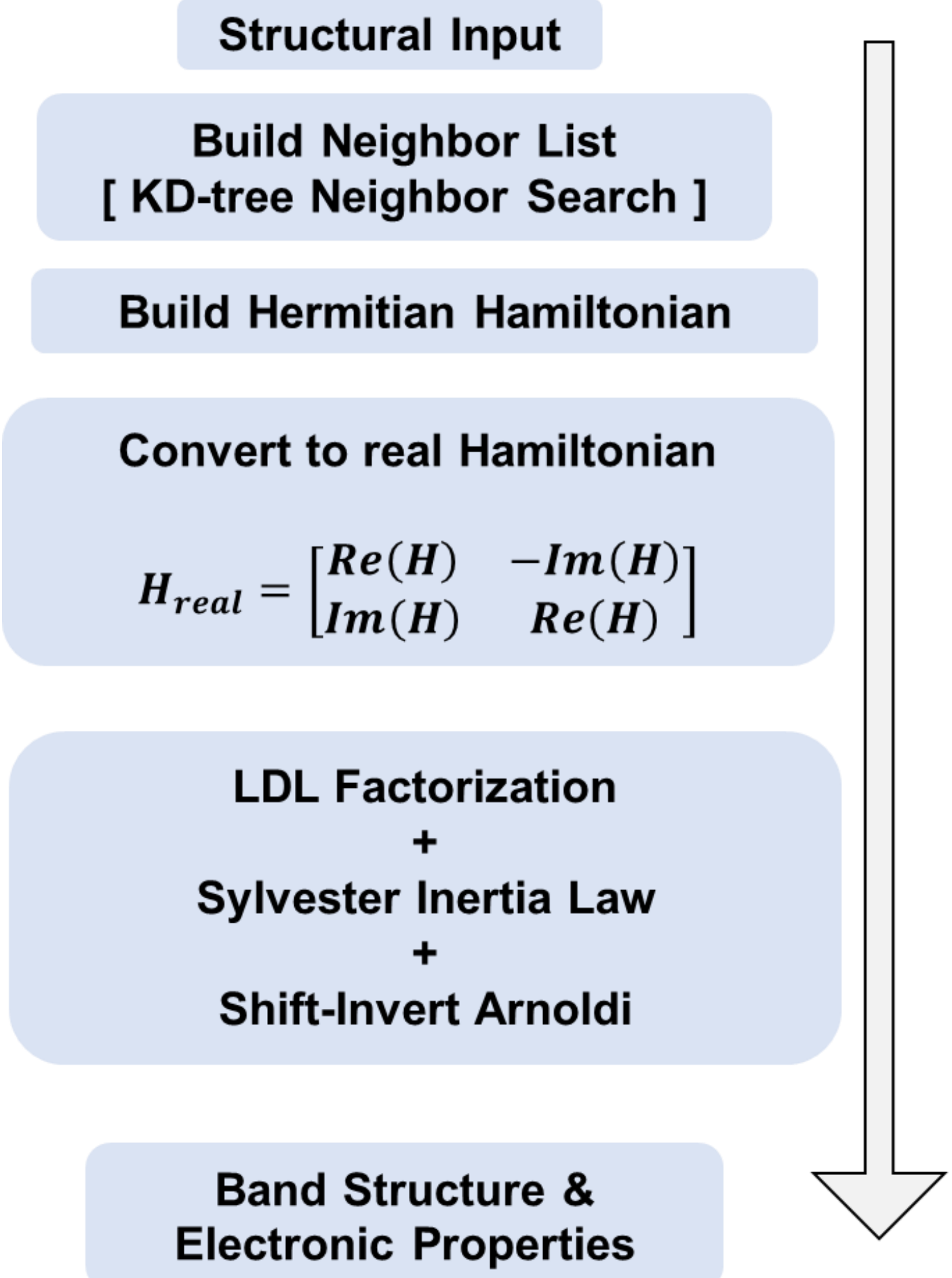

FIG. 1. Schematic workflow of the proposed large-scale tight-binding band-structure algorithm. Starting from a periodic atomic structure, the method constructs a sparse tight-binding Hamiltonian, determines the Fermi level through spectral-window extraction and inertia-based rank calibration, and computes near-Fermi bands along high-symmetry paths.

For a periodic supercell containing $N$ atoms, the tight-binding Hamiltonian $H$ is sparse because only hopping matrix elements $t_{ij}$ between atom pairs within a finite cutoff distance $r_c$ are retained. Computing all pairwise distances directly would scale as $O(N^2)$. By using a KD-tree spatial-indexing algorithm [32] to partition the cell, this step is reduced to $O(N\log N)$. For each atom $i$, we identify all neighboring atoms $j$, including periodic images in adjacent cells, that satisfy $|\boldsymbol{r}_i - \boldsymbol{r}_j| < r_c$. The resulting sparse connectivity pattern is constructed once and reused for all wave vectors $\boldsymbol{k}$. After the neighbor list is established, the sparse Hermitian

Hamiltonian can be assembled efficiently and used in the subsequent eigenvalue calculations.

For TBG, the Bloch Hamiltonian $H(\boldsymbol{k})$ is generally complex Hermitian. While sparse LDL factorization is well-developed for real-symmetric matrices in most high-performance computing libraries [35,36], efficient and robust implementations for complex Hermitian matrices remain challenging. To remove this implementation bottleneck and improve portability across numerical environments, we use an equivalent real-symmetric transformation. Writing $H(\boldsymbol{k}) = A + iB$, where $A$ is real-symmetric and $B$ is real skew-symmetric, we construct

$$H_{\mathrm{real}}(\boldsymbol{k}) = \begin{bmatrix} A & -B \\ B & A \end{bmatrix} \tag{1}$$

As shown in the Methods, the real-symmetric matrix $H_{\mathrm{real}}(\boldsymbol{k})$ duplicates the spectrum of $H(\boldsymbol{k})$. If the ordered eigenvalues of $H(\boldsymbol{k})$ are $\lambda_1, \lambda_2, \ldots, \lambda_N$, then those of $H_{\mathrm{real}}(\boldsymbol{k})$ are $\lambda_1, \lambda_1, \lambda_2, \lambda_2, \ldots, \lambda_N, \lambda_N$. We call $H_{\mathrm{real}}(\boldsymbol{k})$ as the real-symmetric embedding of the original complex $H(\boldsymbol{k})$. Therefore, the j-th eigenvalue of $H(\boldsymbol{k})$ is ranked 2j-1 and 2j in the embedded spectrum of $H_{\mathrm{real}}(\boldsymbol{k})$. This relation is crucial for calibrating energy ranks (e.g., HOMO/LUMO) without dense diagonalization. It preserves the spectral information required for the Fermi-level search. Its main practical advantage is that the rank-calibration problem is transformed into a real-symmetric sparse factorization problem, for which highly optimized real-sparse-LDL solvers are widely available in scientific computing environments. This transformation therefore improves the portability of our algorithm across numerical platforms.

For example, in a charge-neutral graphene system with $N$ atoms and one $p_z$ orbital per carbon atom [37], the number of occupied single-particle states is $N_{occ} = N/2$ [38]. In electron- or hole-doped systems, this count becomes $N_{\mathrm{occ}} = (N - Q)/2$, where $Q$ is the net charge, taken as positive for holes and negative for electrons. The HOMO and LUMO are then identified as the eigenvalues with global ranks $N_{occ}$ and $N_{occ} + 1$, respectively, in the original complex Hamiltonian $H(\boldsymbol{k})$ (typically at the $\boldsymbol{K}$ point for TBG, which is also known as Dirac point [4]). We define $E_F$ as the midpoint between these two eigenvalues.

To determine the ranking of eigenvalues from the embedded spectrum of the real-symmetric matrix $H_{\mathrm{real}}(\boldsymbol{k})$, we combine a shift-invert eigensolver with inertia-based rank calibration. The procedure starts from a trial shift $\sigma$ near the expected Fermi energy (for TBG we typically use $\sigma = 0.78eV$). A small cluster of eigenvalues around $\sigma$ is then computed using the shift-invert Arnoldi method. However, this eigensolver returns only the eigenvalues within a local spectral window and does not provide their global ordering to anchor the actual Fermi energy. The remaining task is therefore to determine the global rank of computed spectral window within the full energy spectrum. For example, we can choose to determine the global rank of the lower edge $E_{min}$ within the computed spectral window of $H_{\mathrm{real}}(\boldsymbol{k})$.

We then use LDL decomposition together with Sylvester's law of inertia [21,39,40]. To avoid possible singular or degenerate cases at $E_{min}$, we choose a reference energy $E_{ref}$ slightly below the lower edge $E_{min}$. Critically, Sylvester's law of inertia relates the number of eigenvalues below $E_{ref}$ in the spectrum of $H_{real}(\boldsymbol{k})$. The sparse LDL factorization of the real-symmetric matrix $H_{real}(\boldsymbol{k}) - E_{ref} I_{2n\times 2n}$ results in a block-diagonal matrix $D$ contains $1 \times 1$ and $2 \times 2$ diagonal blocks. These little diagonal blocks are easy to solve, key to our fast rank-determination. From Sylvester's law of inertia, the accumulated number of negative eigenvalues from these $1 \times 1$ and $2 \times 2$ diagonal blocks equals the number of eigenvalues below $E_{ref}$ in the embedded spectrum of $H_{real}(\boldsymbol{k})$, thereby anchoring the spectral window to the global eigenvalue ranking. The target HOMO and LUMO ranks in the embedded spectrum are $2N_{occ}$ and $2(N_{occ}+1)$, respectively. If both target ranks lie within the current spectral window, the corresponding eigenvalues are selected directly. Otherwise, the trial shift $\sigma$ is moved upward or downward according to the relative position of the current spectral window and the target ranks, and the procedure is repeated until the target ranks are captured.

Once $E_F$ is determined, the near-Fermi band structure is computed along the chosen high-symmetry path. The real-symmetric embedding in $H_{real}(\boldsymbol{k})$ is used only for inertia-based rank calibration at the reference point to obtain $E_F$; the subsequent band calculations are performed using the original complex $H(\boldsymbol{k})$. For each wave vector $\boldsymbol{k}$, $H(\boldsymbol{k})$ is assembled by multiplying the precomputed hopping amplitudes by the Bloch phase factor $\mathrm{e}^{i\boldsymbol{k}\cdot\boldsymbol{d}}$, where $\boldsymbol{d}$ is the interatomic hopping vector including periodic images. We then apply the shift-invert Arnoldi eigensolver to the original $H(\boldsymbol{k})$ with the shift set to $\sigma = E_F$, extracting a fixed number of eigenvalues closest to the Fermi level.

For two-dimensional highly-sparse periodic Hamiltonians, the dominant operations in both the LDL-inertia Fermi-level search and the shift-invert Arnoldi band extraction are sparse factorizations. As shown below, these operations lead to an empirical computational complexity close to $O(N)$, representing a substantial improvement over the conventional $O(N^3)$ bottleneck associated with dense full diagonalization [14–17].

## Ultra-low twist-angle moiré physics of bilayer graphene

Our method is applied to explore a previously inaccessible regime of twisted bilayer graphene (TBG); complete computational details are provided in Methods. A commensurate TBG structure is commonly specified by a pair of integers $(m, n)$ that define the moiré unit cell [41,42]. Here we focus on configurations satisfying $n = m + 1$, and adopt the tight-binding parameters from Moon and Koshino [42]. To account for mechanical strain relaxation, the atomic structures are optimized using LAMMPS [43–45]. Relaxation is considered complete only when the characteristic domain reconstruction and moiré-scale structural features become stable and consistent with previous studies [19,46,47]. In addition to band energies, we compute the local density of states (LDOS) to characterize the real-space distribution of low-energy electronic states. For all LDOS maps, the target energy is fixed at the Fermi level. The resulting site-resolved LDOS is then projected onto the relaxed moiré unit cell to reveal the spatial distribution of electronic states at the Fermi level. To further demonstrate the applicability and robustness of the framework, we explore the effect of uniform perpendicular magnetic fields with intensity approaching practical lab settings, and the effect of dilute isolated topological defects, in ultra-low twisted bilayer graphene with large moiré wavelength.

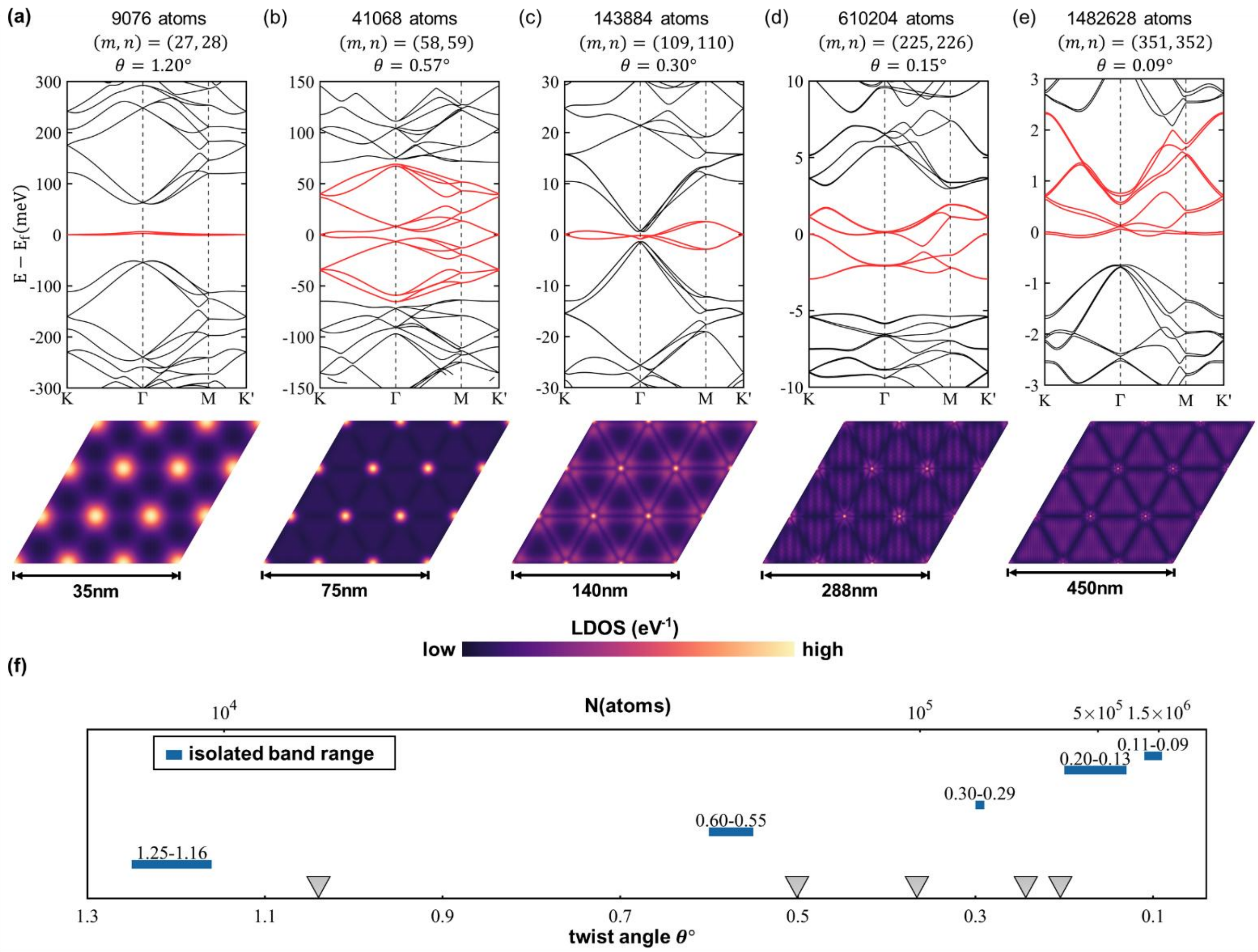

FIG. 2. Low-energy electronic structures, Fermi-level LDOS maps, and isolated-band twist-angle windows of relaxed twisted bilayer graphene. (a-e) Low-energy bands of representative commensurate structures $(m, n) = (27,28)$, $(58, 59)$, $(109,110)$, $(225,226)$ and $(351,352)$, corresponding to $\theta = 1.20°$, $0.57°$, $0.30°$, $0.15°$ and $0.09°$, respectively. The numbers of atoms in the moiré supercells are labeled above each panel. Red curves highlight the isolated low-energy bands near the Fermi level. The LDOS maps below the band structures are evaluated at the Fermi level and projected onto the corresponding relaxed moiré unit cells; the color scale denotes the LDOS intensity. The scale bar indicates the physical length scale of the LDOS map. (f) Twist-angle windows in which isolated low-energy bands are retained, identified from a systematic sequence of commensurate moiré supercells down to 0.09°. Blue horizontal bars mark the isolated-band windows, while gray inverted triangles indicate previously proposed theoretical higher-order magic angles. The upper horizontal axis gives the approximate number of atoms in the corresponding commensurate supercells.

### *Electronic structure at ultra-low twist angles*

We computed a systematic sequence of commensurate relaxed moiré supercells down to 0.09°. Panels in Figs. 2(a-e) summarize the low-energy band structures and Fermi-level LDOS maps of relaxed TBG from the first magic-angle regime to the ultra-low-angle limit. The systems range from $(m,n) = (27,28)$, containing 9076 atoms, to $(m,n) = (351,352)$, containing 1482628 atoms. All structures were fully relaxed before the band and LDOS calculations [20,47].

The first magic-angle structure is identified by the strong suppression of the low-energy bandwidth (bandwidth below 16meV), consistent with the conventional criterion [48]. Figure 2(a) shows the representative structure $(m,n) = (27,28)$ with the minimum low-energy bandwidth (7meV). At lower twist angles, the highlighted low-energy bands are energetically separated from nearby remote bands but retain finite and momentum-dependent dispersion. Pronounced flat-band behavior is therefore limited to the first magic-angle regime. The separation of the low-energy bands suppresses hybridization with higher-energy states and enhances the low-energy spectral weight, providing favorable conditions for interaction-driven electronic behavior.

Figure 2(f) summarizes the twist-angle windows in which such isolated low-energy band clusters are retained. These windows occur over finite angle ranges, from $1.25° - 1.06°$, $0.60° - 0.55°$, $0.30° - 0.29°$, $0.20° - 0.13°$ to $0.11° - 0.09°$. There are some noticeable discrepancies between them and previously proposed ideal continuum-model higher-order magic-angles [1], as shown in gray inverted triangles along the horizontal axis. The offsets likely reflect the combined influence of lattice relaxation [47] and tight-binding parameterization [42]. Importantly, the finite width of each isolated-band window indicates a finite angular tolerance: isolated low-energy band features persist over a range of twist angles rather than appearing only at a single precisely tuned angle. This interpretation is also consistent with nanoscale experimental studies of magic-angle TBG, which reported that magic-angle transport features can coexist with local twist-angle disorder of about 0.1° [49].

The Fermi-level LDOS maps in Fig. 2(a-e) further show that the near-Fermi states remain spatially inhomogeneous within the relaxed moiré unit cell, with enhanced spectral weight around the AA regions. Across the investigated twist-angle range, the LDOS retains the moiré-scale inhomogeneity while its spatial localization evolves with the moiré length scale. Together, the band structures and Fermi-level LDOS maps demonstrate that the proposed method can resolve not only the low-energy spectrum but also the real-space character of electronic states in TBG systems containing more than 1.4 million atoms.

***Effect of magnetic fields***

We next investigate the response of the low-energy electronic states to a uniform magnetic field applied perpendicular to the graphene layers. This setting is important for ultra-low-angle TBG because in a periodic Peierls-phase treatment, the magnetic flux through the moiré unit cell must be quantized as $\boldsymbol{B} \cdot (\boldsymbol{T}_1 \times \boldsymbol{T}_2) = q\Phi_0$, where $\Phi_0 = h/e$ and $\boldsymbol{T_1}, \boldsymbol{T_2}$ are lattice vectors of the original moiré supercell [50]. Consequently, the minimum allowed perpendicular magnetic field is inversely proportional to the moiré unit-cell area. For magic-angle TBG near $\theta = 1.08°$, the first nonzero perpendicular field is around 28 T [51,52], already beyond typical laboratory fields [49,51]. In contrast, ultra-low-angle moiré supercells greatly enlarge the unit-cell area, reduce the corresponding magnetic field strength to the tesla range used in low-field Landau-level measurements [49].

The magnetic field is introduced through the Peierls substitution, in which each hopping matrix element is multiplied by a phase factor, $t_{ij} \rightarrow t_{ij}^{B} = t_{ij}\mathrm{e}^{i\phi_{ij}}$. A general gauge choice may break the translational periodicity of the Hamiltonian, so we adopt the periodic Peierls-phase construction for atomistic two-dimensional systems [50,51] , under which the allowed perpendicular field is set by the flux through the moiré unit cell. Figure 3(a) shows the magnetic field corresponding to one flux quantum per moiré unit cell, $q = 1$, as a function of twist angle. As the twist angle approaches $0.1°$, the $q = 1$ field decreases to about $0.21$ T; even for $q = 5$, the corresponding field is only about $1$ T.

The low-energy magnetic band structures in Figs. 3(b–f) are calculated for the same commensurate relaxed TBG structures as those in Fig. 2, with the magnetic flux fixed at $q = 5$ in each moiré unit cell. The corresponding magnetic field decreases from $B = 173.98\,T$ at $\theta = 1.20°$ to $B = 1.07\,T$ at $\theta = 0.09°$. Compared with the zero-field band structures in Fig. 2(a), the perpendicular magnetic field strongly suppresses the dispersion of the low-energy bands. The resulting bands become much flatter and more clearly separated from the remote dispersive bands, forming isolated groups of minibands near the Fermi level. This behavior is consistent with the formation of Landau-level-like states in a periodic moiré superlattice under quantized magnetic flux [50,51]. Similar isolated flat-band features are also observed in ultra-low-angle structures, even though the required magnetic fields are much smaller.

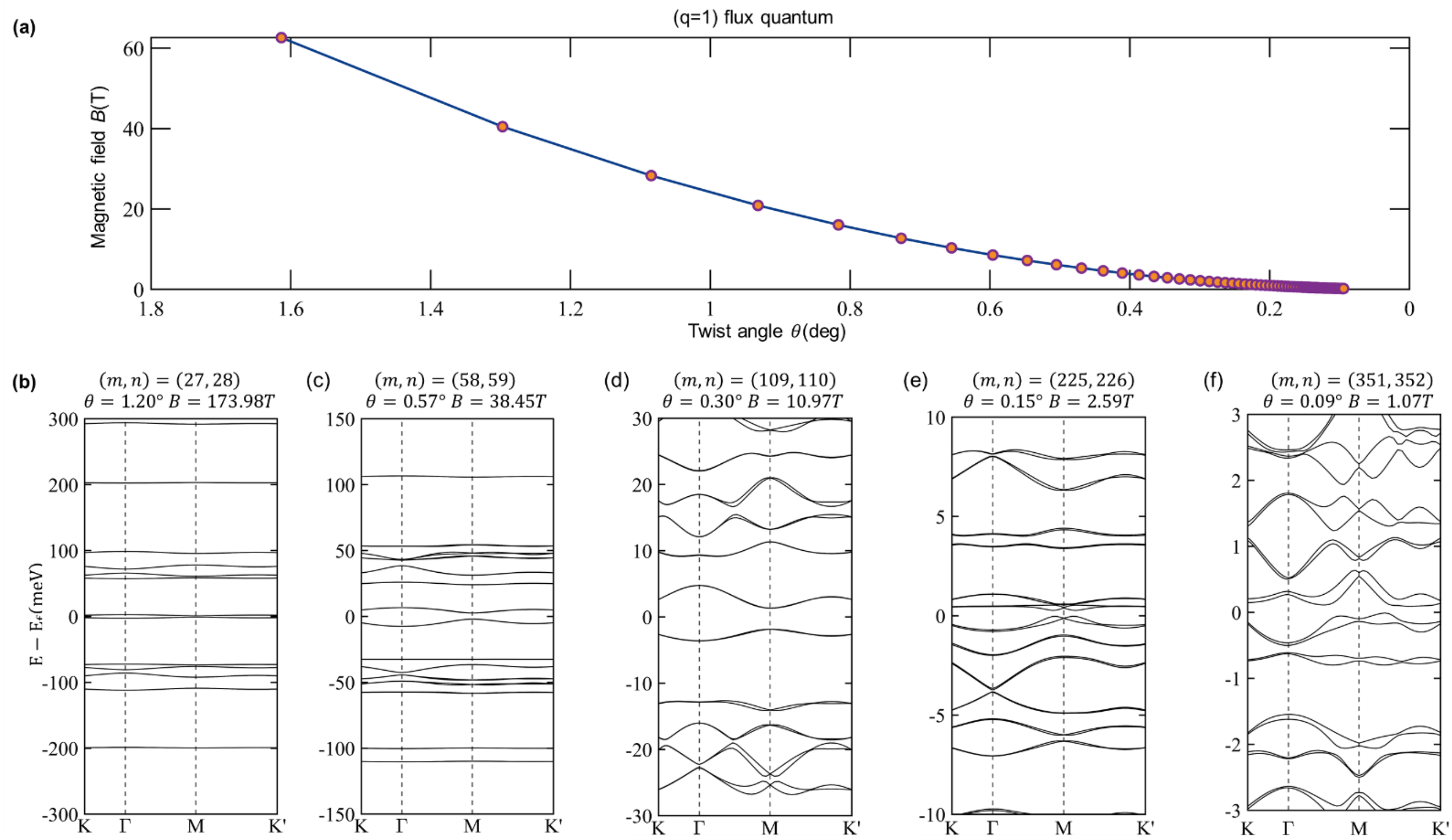

FIG. 3. Perpendicular magnetic fields and low-energy magnetic band structures of relaxed twisted bilayer graphene. (a) Magnetic field corresponding to one flux quantum per moiré unit cell, $q = 1$, as a function of twist angle. (b-f) Low-energy magnetic band structures calculated with the magnetic flux through each moiré unit cell fixed at $q\Phi_0$, with $q = 5$.

### *Effect of dilute Stone-Wales defects*

We further examine the effect of dilute Stone-Wales defects in magic-angle TBG. Recent progress in graphene growth, transfer, and encapsulation has made it possible to prepare high-quality graphene with very weak defect signatures [53–55]. In this low-defect regime, the relevant question becomes how a rare local lattice defect perturbs the low-energy electronic structure of TBG. Among intrinsic structural defects, the Stone-Wales (SW) defect provides a minimal model for experimentally accessible topological lattice disorder: it converts four neighboring hexagons into two pentagons and two heptagons without changing the total number of carbon atoms, and can be generated under non-equilibrium synthesis or irradiation conditions [56,57].

Previous studies have shown that SW defects introduce concentration- and supercell-size-dependent electronic signatures in graphene, including defect-related bands and Dirac-cone shifts [57], energy-dependent band broadening without strong Dirac-point resonance [58], as well as defect-induced flat bands accompanied by van Hove singularities in TBG [59]. To model the dilute-defect limit under periodic boundary conditions, we introduce one SW defect into $n \times n$ replicated supercells of the fully relaxed magic-angle structure at $\theta = 1.08°$, with $n = 2, 5,$ and $8$ [Fig. 4(a)]. This setup corresponds to a periodic array of SW defects, where increasing $n$ increases the defect-defect separation and reduces the effective defect concentration [57,59].

For each supercell, we place one SW defect in AA, AB, and saddle-point (SP) regions, followed by full structural relaxation. The corresponding defect-induced moiré reconstruction maps are provided in the Supplementary Data. Fig. 4(b–d) compares the near-$E_F$ band structures of replicated TBG supercells with (red) and without (black) SW defect. In the 2×2 supercell, where the effective defect concentration is relatively high, the SW defect visibly modifies the low-energy bands regardless of its location. As the supercell size increases and the effective defect concentration decreases, the effect of the defect becomes strongly location dependent: defects near AA region still produce visible low-energy band changes, whereas defects in AB and SP regions have only weak influence. This location dependence is consistent with the real-space LDOS distribution shown in Fig. 2. Since the low-energy states are mainly concentrated near AA regions, a defect placed near an AA region perturbs the low-energy bands more strongly. These results suggest that, in dilute defective TBG samples, the band-structure response can provide a useful fingerprint for identifying the defect location.

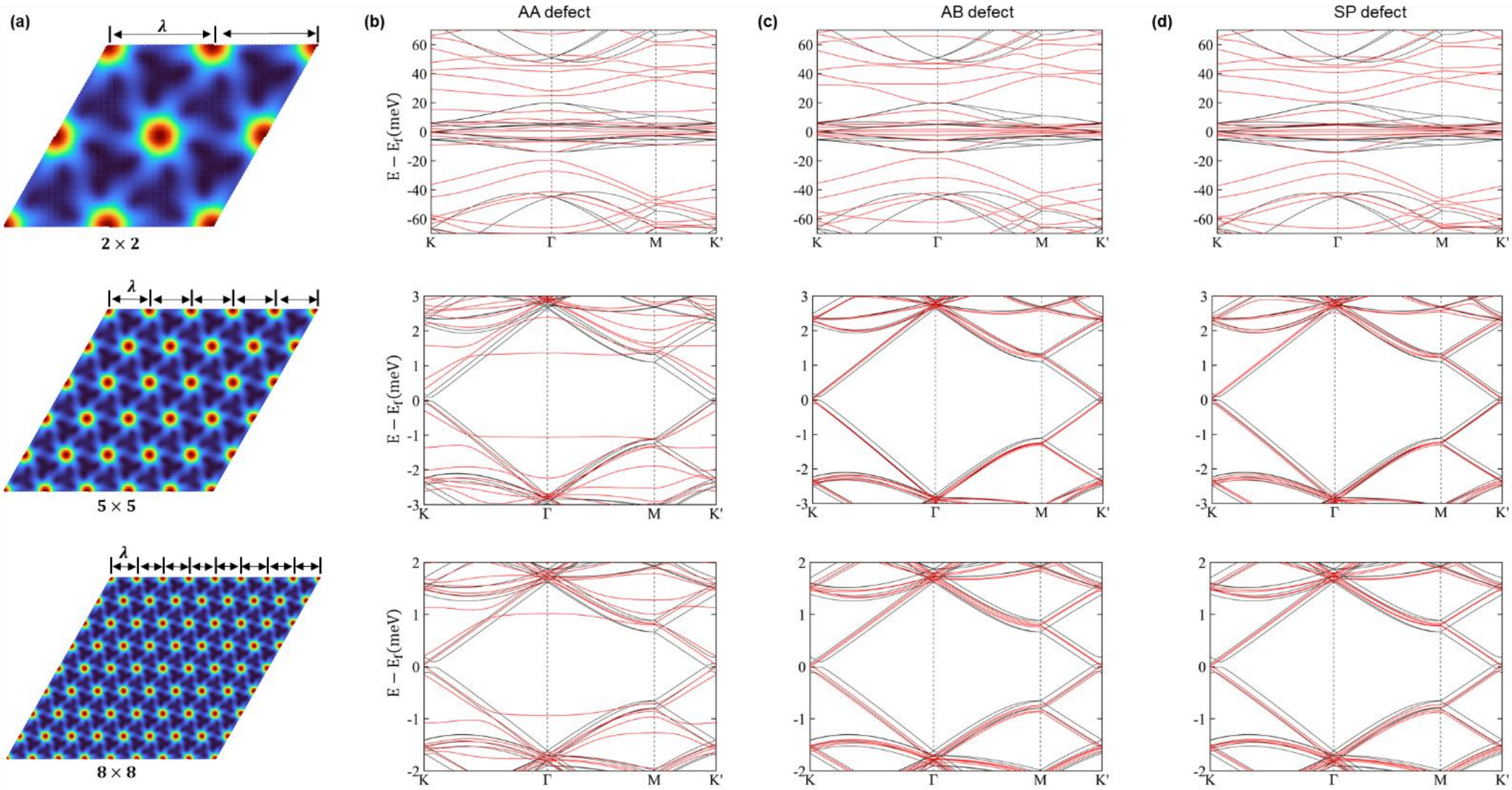

FIG. 4. Effect of a single Stone-Wales defect on replicated magic-angle TBG supercells. Rows show 2×2, 5×5, and 8×8 supercells constructed from the relaxed $\theta = 1.08^\circ$ moiré unit cell. (a) Enlarged moiré supercells colored by the top-layer $z$ coordinate. (b-d) Near-$E_F$ band structures after introducing one SW defect in AA, AB, and SP regions, respectively. Black curves show the fully relaxed pristine supercell bands, and red curves show the corresponding relaxed SW-defective bands.

## Scaling performance

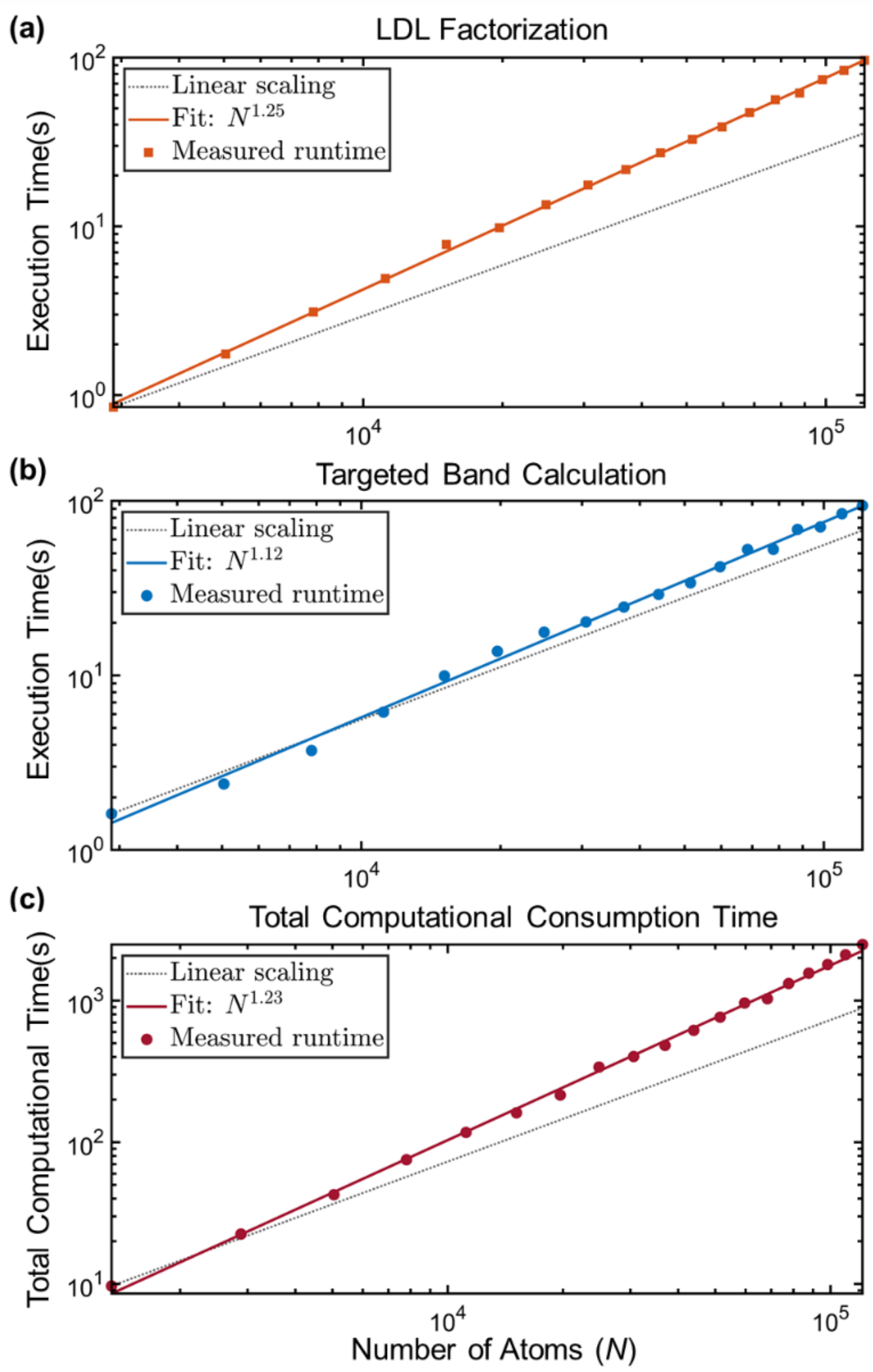

FIG. 5. Near-linear scaling performance of the proposed method for twisted bilayer graphene systems with different numbers of atoms. (a) Runtime for LDL factorization of the transformed real-symmetric Hamiltonian. (b) Runtime of the targeted eigensolver for one $\boldsymbol{k}$-point. (c) Total runtime for computing a complete near-Fermi band structure.

To assess the scaling performance of our algorithm, we construct a series of TBG moiré unit cells by systematically varying the twist angle specified by the integer pair $(m, n)$. Figure 5 shows the measured runtime for obtaining 40 bands near the Fermi level. We separately evaluate the two main computational steps: LDL factorization of the $2N \times 2N$ transformed real-symmetric Hamiltonian used in the Fermi-level search [Fig. 5(a)], and a single targeted eigenvalue calculation at one $\boldsymbol{k}$ point using the shift-invert eigensolver [Fig. 5(b)]. The measured runtimes follow empirical power-law scaling of approximately $N^{1.25}$ for LDL factorization and $N^{1.12}$ for a single targeted eigensolver run. The total runtime required to compute a complete near-Fermi band structure is shown in Fig. 5(c), with an empirical scaling of approximately $N^{1.23}$. The fitted exponents indicate near-linear scaling over the tested system-size range, while deviations from ideal linear behavior may reflect implementation details and variations in the sparsity patterns of TBG Hamiltonians. Overall, the full workflow remains substantially more efficient than conventional dense diagonalization with $O(N^3)$ complexity. The near-Fermi band structures of million-atom TBG supercells can be obtained within days on a single workstation, rather than requiring large-scale supercomputing resources.

## Conclusion

This work presents a scalable computational framework for large-scale electronic-structure calculations. Near-linear scaling is achieved by transforming the Hamiltonian into an equivalent real-symmetric form and combining LDL decomposition with Sylvester's law of inertia for inertia-based spectral slicing. This approach enables accurate band-structure calculations for systems containing more than one million atoms, thereby bridging the scale gap between first-principles accuracy and mesoscopic sample dimensions.

Our method provides a scalable computational platform for quantum materials research. High-fidelity local electronic parameters obtained from small-scale density functional theory (DFT) calculations can be directly incorporated into this scalable framework, allowing systematic exploration of macroscopic quantum phenomena, such as the evolution of flat bands and soliton networks in ultra-low-angle twisted bilayer graphene, that were previously beyond the reach of both ab initio and conventional tight-binding approaches. Furthermore, the algorithm can provide large-scale electronic-structure datasets for training and validating next-generation data-driven models. By extending quantum-accurate simulations to experimentally relevant length scales, this work opens a practical route toward data-driven discovery in complex materials.

## Methods

### ***Details of tight-binding calculations***

The tight-binding Hamiltonian is written as

$$H = \sum_{i,j} t\left(\boldsymbol{R}_i - \boldsymbol{R}_j\right)\left|\boldsymbol{R}_i\right\rangle\left\langle\boldsymbol{R}_j\right| + H.c., \tag{2}$$

where $\boldsymbol{R}_i$ denotes the atomic coordinate, and $|\boldsymbol{R}_i\rangle$ is the wavefunction localized at site $i$. The term $t\left(\boldsymbol{R}_i - \boldsymbol{R}_j\right)$ is the transfer integral between atoms $i$ and $j$. Following the standard Slater–Koster formula [41] ,

$$t(\boldsymbol{d}) = V_{pp\pi}(d)\left[1 - \left(\frac{\boldsymbol{d}\cdot\boldsymbol{e}_z}{d}\right)^2\right] + V_{pp\sigma}(d)\left(\frac{\boldsymbol{d}\cdot\boldsymbol{e}_z}{d}\right)^2,$$

$$V_{pp\pi}(d) = V_{pp\pi}^0 e^{\left(-\frac{d-a_0}{\delta}\right)}, \tag{3}$$

$$V_{pp\sigma}(d) = V_{pp\sigma}^0 e^{\left(-\frac{d-d_0}{\delta}\right)}.$$

Here $\boldsymbol{d} = \boldsymbol{R}_i - \boldsymbol{R}_j$ is interatomic hopping vector (including periodic images), and $\boldsymbol{e}_z$ is the unit vector along the *z*-axis. From Moon and Koshino [42], we use $V_{pp\pi}^0 = -2.7\ eV, V_{pp\sigma}^0 = 0.48\ eV,\ d_0 = 0.335\ nm,$ and $\delta = 0.184\sqrt{3}a_0$ ( $a_0 = 0.142\ nm$ is carbon-carbon bond length). The transfer integral is neglected for $d > 4a_0$.

The atomic structures are relaxed using LAMMPS [44]. Intra-layer interactions are described by the REBO potential [45], while the interlayer interactions are modeled using the registry-dependent interlayer potential (ILP) [43]. Energy minimization is performed with the conjugate-gradient algorithm until the relative change in total energy between successive iterations falls below $10^{-13}$. To ensure that this convergence criterion corresponds to physically relaxed structures, we further examine the resulting moiré patterns and reconstructed atomic configurations until they become stable and consistent with previous studies of fully relaxed supercell [19,46,47].

In addition to the band energies, we compute the local density of states (LDOS) to characterize the real-space distribution of low-energy electronic states. For Bloch eigenstates $\psi_{n\boldsymbol{k}}$, the site-resolved LDOS is defined as,

$$\rho_i(E) = \frac{1}{N_k} \sum_{\boldsymbol{k},n} |\psi_{n\boldsymbol{k}}(i)|^2 g_\eta(E - E_{n\boldsymbol{k}}), \tag{4}$$

where $\psi_{n\boldsymbol{k}}(i)$ is the wave-function amplitude on atomic site $i, E_{n\boldsymbol{k}}$ is the corresponding eigenvalue, $N_k$ is the number of sampled wave vectors, and $g_\eta$ is a Gaussian broadening function $g_\eta(E - E_{n\boldsymbol{k}}) = \frac{1}{\eta\sqrt{2\pi}} e^{-\frac{(E-E_{n\boldsymbol{k}})^2}{2\eta^2}}$ with broadening $\eta = 0.005eV$.

For the LDOS maps, the target energy is fixed at the Fermi level. After shifting the eigenvalues by $E_F$, namely $\tilde{E}_{n\boldsymbol{k}} = E_{n\boldsymbol{k}} - E_F$, the Fermi-level LDOS is evaluated as

$$\rho_i\left(\tilde{E}_F\right) = \frac{1}{N_k}\sum_{\boldsymbol{k},n} |\psi_{n\boldsymbol{k}}(i)|^2 \frac{1}{\eta\sqrt{2\pi}} e^{-\frac{\tilde{E}_{n\boldsymbol{k}}^2}{2\eta^2}} \tag{5}$$

As shown in the lower panel of Fig. 2, the LDOS maps are projected onto the relaxed moiré unit cell to show the spatial distribution of electronic states at the Fermi level.

### *Real-symmetric Transformation and Inertia Relation*

Let $H(\boldsymbol{k})$ be an $N \times N$ complex Hermitian matrix and decompose it into real and imaginary parts as

$$H(\boldsymbol{k}) = A + iB, \tag{6}$$

where $A$ is real-symmetric ($A^\top = A$) and $B$ is real skew-symmetric ($B^\top = -B$). Suppose that $\lambda$ is an eigenvalue of $H(\boldsymbol{k})$ with eigenvector $\boldsymbol{x} = \boldsymbol{u} + i\boldsymbol{v} \in \mathrm{C}^N$ ($\boldsymbol{u}, \boldsymbol{v} \in \mathbf{R}^N$). From the definition of eigenvector,

$$H(\boldsymbol{k})\boldsymbol{x} = \lambda\boldsymbol{x}. \tag{7}$$

Separating the real and imaginary parts gives

$$\begin{cases} A\boldsymbol{u} - B\boldsymbol{v} = \lambda\boldsymbol{u} \\ B\boldsymbol{u} + A\boldsymbol{v} = \lambda\boldsymbol{v} \end{cases}. \tag{8}$$

These two equations can be written in block form as

$$\begin{pmatrix} A & -B \\ B & A \end{pmatrix}\begin{pmatrix} \boldsymbol{u} \\ \boldsymbol{v} \end{pmatrix} = \lambda\begin{pmatrix} \boldsymbol{u} \\ \boldsymbol{v} \end{pmatrix} \tag{9}$$

Thus, the real-symmetric block matrix

$$H_{real}(\boldsymbol{k}) = \begin{bmatrix} A & -B \\ B & A \end{bmatrix} \tag{10}$$

has $(\boldsymbol{u}, \boldsymbol{v})^{\top}$ as a real eigenvector with eigenvalue $\lambda$. Multiplying the original complex eigenvector by $i$ yields $i\boldsymbol{x} = -\boldsymbol{v} + i\boldsymbol{u}$, which corresponds to another real eigenvector $(-\boldsymbol{v}, \boldsymbol{u})^{\top}$ with the same eigenvalue. Therefore, each eigenvalue of $H(\boldsymbol{k})$ appears twice in $H_{real}(\boldsymbol{k})$.

Since $H(\boldsymbol{k})$ is Hermitian, its eigenvectors form a complete basis. Applying the above construction to the full eigenbasis shows that the spectrum of $H_{real}(\boldsymbol{k})$ consists of the spectrum of $H(\boldsymbol{k})$ repeated twice, counting multiplicity. In other words, if the ordered eigenvalues of $H(\boldsymbol{k})$ are $\lambda_1 \le \lambda_2 \le \cdots \le \lambda_N$, then those of $H_{real}(\boldsymbol{k})$ are $\lambda_1, \lambda_1, \lambda_2, \lambda_2, \dots, \lambda_N, \lambda_N$. Therefore, the j-th eigenvalue of $H(\boldsymbol{k})$ appears at ranks 2j-1 and 2j in $H_{real}(\boldsymbol{k})$.

This spectral duplication is preserved under a scalar energy shift. For any reference energy $E_{\mathrm{ref}}$, the shifted matrix $H_{real}(\boldsymbol{k}) - E_{ref} I_{2N}$ is exactly the real-symmetric embedding of $H(\boldsymbol{k}) - E_{ref} I_N$. Therefore, the same rank relation holds for the shifted matrices. In the Fermi-level search, the HOMO eigenvalue $\lambda_{\mathrm{N}_{occ}}$ appears at ranks $2N_{occ} - 1$ and $2N_{occ}$, whereas the LUMO eigenvalue $\lambda_{\mathrm{N}_{occ+1}}$ appears at ranks $2N_{occ} + 1$ and $2(N_{occ} + 1)$. In our implementation, we use the second copy of each duplicated pair, giving the target ranks $2N_{occ}$ and $2(N_{occ} + 1)$.

The same relation can be expressed in terms of inertia, which is the quantity used in the LDL-based rank calibration. Counting eigenvalues below $E_{\mathrm{ref}}$ is equivalent to counting negative eigenvalues of the shifted matrix. Therefore, the negative inertia of $H_{real}(\boldsymbol{k}) - E_{ref}I_{2N}$ is twice the negative inertia of $H(\boldsymbol{k}) - E_{ref}I_N$. This means that the shifted real-symmetric matrix gives the global rank count in the embedded spectrum, equal to twice the corresponding count for the original complex Hermitian Hamiltonian. This relation justifies the real-symmetric LDL-inertia procedure used in the main text to locate the HOMO/LUMO window.

## Supplementary Data

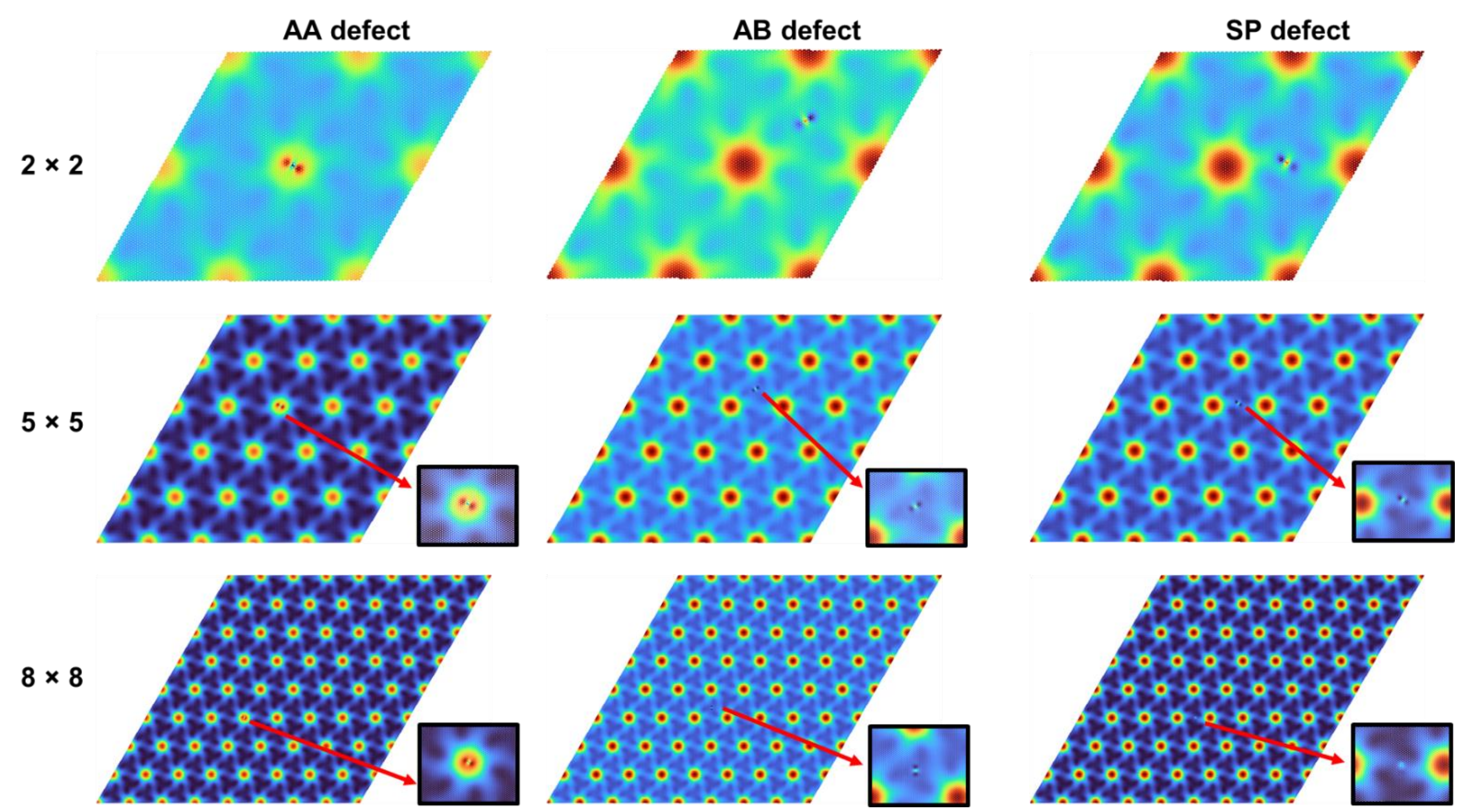

**Fig. S1**. Relaxed top-layer $z$-coordinate maps of replicated magic-angle TBG supercells containing one Stone-Wales defect. The rows show 2 × 2, 5 × 5, and 8 × 8 replicated supercells constructed from the relaxed $\theta = 1.08°$ moiré unit cell. The columns indicate defects placed in AA, AB, and saddle-point (SP) regions. For the 5 × 5 and 8 × 8 supercells, magnified insets show the local structural distortion induced by the defect. In the 2 × 2 supercells, the moiré-scale reconstruction remains visible after introducing a single Stone-Wales defect. As the replicated supercell size increases and the effective defect concentration decreases, the local defect has a weaker influence on the global moiré pattern.